\documentclass[aps,twocolumn,groupedaddress]{revtex4}  
\usepackage{booktabs}

\usepackage{amssymb}
\usepackage{amsmath}
\usepackage{amssymb}
\usepackage{graphicx}
\usepackage{subfigure}
\usepackage{color}
\usepackage{mathrsfs}
\usepackage[dvipsnames]{xcolor}
\DeclareMathOperator{\sech}{sech}
\DeclareMathOperator{\arctanh}{arctanh}

\usepackage[breaklinks=true,colorlinks=true]{hyperref}
\hypersetup{colorlinks=true,citecolor=blue,linkcolor=blue,urlcolor=blue}

\usepackage[utf8]{inputenc}

\begin{document}
\immediate\write16{<<WARNING: LINEDRAW macros work with emTeX-dvivers
                    and other drivers supporting emTeX \special's
                    (dviscr, dvihplj, dvidot, dvips, dviwin, etc.) >>}

\title{Fermion bound states from Yukawa coupling with periodic bosonic background}

\author{Dionisio Bazeia$^{1}$ and Fabiano C. Simas$^{2,3}$}

\email{bazeia@fisica.ufpb.br; fc.simas@ufma.br}

\affiliation{$^{1}$Departamento de F\'isica, Universidade Federal da Para\'iba, 58051-970 Jo\~ao Pessoa, PB, Brazil\\
$^{2}$Departamento de Física, Universidade Federal do Maranhão (UFMA), Campus Universitário do Bacanga, 65085-580, São Luís, Maranhão, Brazil\\
$^{3}$Programa de P\'os-Gradua\c c\~ao em F\'isica, Universidade Federal do Maranh\~ao (UFMA),\\ Campus Universit\'ario do Bacanga, 65085-580, S\~ao Lu\'is, Maranh\~ao, Brazil}

\begin{abstract}

The Yukawa coupling of fermions with a periodic bosonic background is shown to give rise to several bound states to the fermionic spectrum, with some bound states gluing together around specific energy eingenvalues as the Yukawa coupling increases. This effect induces the presence of degenerate energy states inside the fermionic gap and may be of current interest.
\end{abstract}

\maketitle

Almost 50 years ago, Jackiw and Rebbi \cite{jackiw} unveiled an interesting effect, the possibility of fermion fractionalization, directly connected with the presence of a zero mode inside the otherwise empty spectrum of bound states. In this case, the fermionic interaction occurs via Yukawa coupling with the scalar background generated by the kinklike configuration of the standard $\phi^4$ model. The result impacted the physics in distinct directions, in particular, in the study of electric conduction in polymers \cite{su,jackiw1,su1}; for a review see, e.g., Ref. \cite{niemi}. 

More recently, in Ref. \cite{baz1} the investigation considered a different model, in which the Dirac Lagrangian now contains a Yukawa coupling with a kinklike configuration that engenders internal structure. In this new case, the presence of kinklike solution that engenders internal structure has contributed to the appearance of several bound states in the spectrum. An interesting feature here is that the internal structure trapped to the kink was shown to appear due to the geometric constriction present in the model \cite{baz2}, which is directly connected with the effects of the geometric constriction previously investigated experimentally in magnetic materials in Ref. \cite{IBM}.

We have learned from Refs. \cite{jackiw,baz1} that the presence of fermionic bound states depends importantly on the specific form of the scalar background, and this has inspired us to study another possibility, recently described in Ref. \cite{baz3}, in which the scalar field is described by a periodic configuration. Toward this goal, let us now present the Lagrangian ${\cal{L}}={\cal{L}}_f+{\cal{L}}_b$, in which ${\cal{L}}_f$ describes the interaction between a massless fermion and a scalar field through the Yukawa coupling
\begin{eqnarray}\label{fermion}
    {\mathcal{L}_f} & = & \frac12 \Bar{\psi}i\gamma^{\mu} \partial_{\mu}\psi - g\phi\Bar{\psi}\psi.
\end{eqnarray}
Here, $\psi$ and  $\phi$ represent the Dirac and the scalar fields, and $g$ is a real parameter that describes the Yukawa interaction. We do not consider backreaction of the fermion field into the scalar structure. The model is defined in $1+1$ spacetime dimensions and we use natural units, with the fields, space and time variables and the Yukawa parameter $g$ being rescaled to become dimensionless quantities. The bosonic portion ${\cal L}_b$ of the Lagrangian ${\cal{L}}={\cal{L}}_f+{\cal{L}}_b$ will be briefly described below, also using dimensionless fields, time, space and parameters. In the original paper by Jackiw and Rebbi \cite{jackiw}, the scalar field was described by 
\begin{eqnarray}
    \phi(x) = v \tanh( x/l),
    \label{phi1}
\end{eqnarray}
which appears from the standard $\phi^4$ model 
\begin{eqnarray}
    {\cal L}_s =\frac12\partial_\mu\phi\partial^\mu\phi-\frac{1}{2l^2}(v^2-\phi^2)^2.
\end{eqnarray}
The solution \eqref{phi1} represents a kinklike configuration, here supposed to have amplitude $v$, real and positive, with $l$ also being a real and positive parameter that controls the width of the localized structure.

In the present work, however, we shall bring novelties considering the fermion field in the presence of the scalar field described by the following Lagrangian density \cite{baz2}
\begin{equation}
{\mathcal{L}_b} =  \frac12 f(\chi) \partial_{\mu} \phi \partial^{\mu} \phi + \frac12 \partial_{\mu} \chi \partial^{\mu} \chi - V(\phi,\chi),
\end{equation}
where $V(\phi,\chi)$ is the potential and $f(\chi)$ is a positive function of the second real scalar field $\chi$ which modifies the kinetic term of the $\phi$ field. For static solutions, the equations of motion for the scalar fields are given by
\begin{eqnarray}
    \frac{d}{dx}\bigg(f(\chi)\frac{d\phi}{dx} \bigg) = \frac{\partial V}{\partial \phi}, \quad \frac{d^2\chi}{dx^2}-\frac12 \frac{df(\chi)}{d\chi}\bigg(\frac{d\phi}{dx} \bigg)^2 = \frac{\partial V}{\partial \chi}.
\end{eqnarray}
Following Refs. \cite{baz1,baz2,baz3}, we suppose the potential has the form
\begin{eqnarray}
    V(\phi,\chi) &=& \frac{W^2_{\phi}}{2f(\chi)} + \frac{W^2_{\chi}}{2},
\end{eqnarray}
where $W=W(\phi,\chi)$ is an auxiliary function and $W_{\phi}=\partial W/\partial \phi$ and $W_{\chi}=\partial W/\partial \chi$. We then consider that $W$ obeys $W_\phi=1-\phi^2$ and $W_\chi=\alpha(1-\chi^2)$, where $\alpha$ is another real and positive parameter, also dimensionless. In this case one gets
\begin{eqnarray}
    V(\phi,\chi) &=& \frac12 \frac{(1-\phi^2)^2}{f(\chi)} + \frac12 \alpha^2(1-\chi^2)^2.
\end{eqnarray}
From this, we can write the following first-order equations $d\phi/dx = (1-\phi^2)/f(\chi)$ and $d\chi/dx = \alpha(1-\chi^2)$, which are responsible for minimizing the energy to the form $E=|W(\phi(+\infty),\chi(+\infty))-W(\phi(-\infty),\chi(-\infty))|$.

As observed in Ref.~\cite{baz2}, the first-order equation for $\chi$ can be solved independently, since it does not depend on $\phi$. Furthermore, following \cite{baz3}, the choice of the function $f(\chi)$ as $1+\cos(\arctanh(\chi))$ leads us to obtain the solution for $\phi$ in the form
\begin{eqnarray}
    \phi(x) = \tanh\bigg(\frac{1}{\alpha}\tan\bigg(\frac{\alpha x}{2} \bigg) \bigg),
    \label{phi2}
\end{eqnarray}
which describes a periodic function, with period $2\pi/\alpha$, to be used as the background scalar field to control the fermionic behavior.

As we have learned from nonrelativistic quantum mechanics \cite{QM}, a periodic and attractive potential may, for instance, under the tight-binding approximation, give rise to a band of energy states. For this reason, we believe it is of current interest to use the above Yukawa coupling to investigate how the system behaves in the relativistic framework. Moreover, in graphene, graphite and in carbon nanotubes, electrons may be seem as Dirac fermions, so the present study may have widen scope; see, e.g., Refs. \cite{A,B,C,D,Rev}. Since the present investigation requires that fermions interact with localized bosonic structures, one will also need the presence of defects in such materials, but this is another subject of current interest, as one can see, for instance, in Refs. \cite{D1,Dd,Ddd,TBG,D2,Rev} and in references therein.

 To investigate the issue under consideration, one considers the Lagrangian \eqref{fermion} to get to the equation motion for the fermion field

\begin{eqnarray}
(i\gamma^{\mu} \partial_{\mu} - 2g\phi)\psi = 0. 
\label{eq_fermion}
\end{eqnarray}
We use the following representation of the Dirac gamma matrices: $\gamma^0=\sigma_1$, $\gamma^1=i\sigma_3$ and $\gamma^5=\sigma_2$. Additionally, since the scalar field does not depend on time, we suppose that the fermion field has the form

\begin{eqnarray}
\psi(x,t)=e^{-iEt} \begin{pmatrix} \psi_+(x) \\ \psi_-(x)\end{pmatrix}.
\end{eqnarray}
Thus, the equations of motion for the fermion components are given by

\begin{eqnarray}\label{DE1}
       \bigg( \partial_x - 2g \phi \bigg) \psi_- & = & -E\psi_+,\\
       \bigg( \partial_x + 2g \phi \bigg) \psi_+ & = & E\psi_-.\label{DE2}
\end{eqnarray}
Setting $E=0$, we can thus derive the fermion zero mode

\begin{eqnarray}
    \psi_0(x)=C \begin{pmatrix} e^{-2g\int^x \phi(y) dy'} \\ 0 \end{pmatrix},
\end{eqnarray}
where $C$ stands for the normalization factor. It depends on $g$ and $\alpha$, and can be calculated numerically. For instance, for $\alpha=0.5$ and $g=0.4$, we have $C=0.579$.

\begin{figure}[!ht]
\begin{center}
  \centering
   \includegraphics[width=0.45 \textwidth]{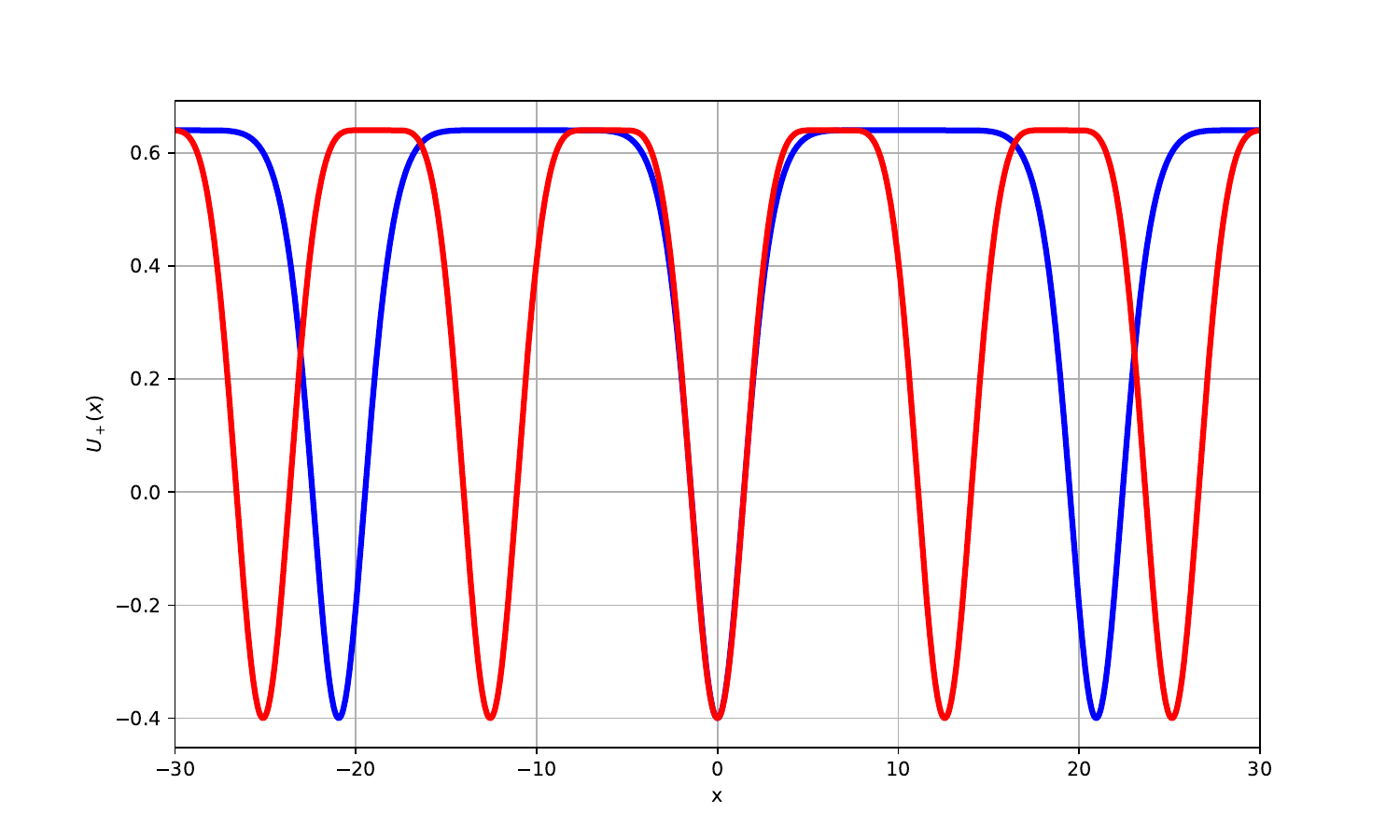}
    \includegraphics[width=0.45 \textwidth]{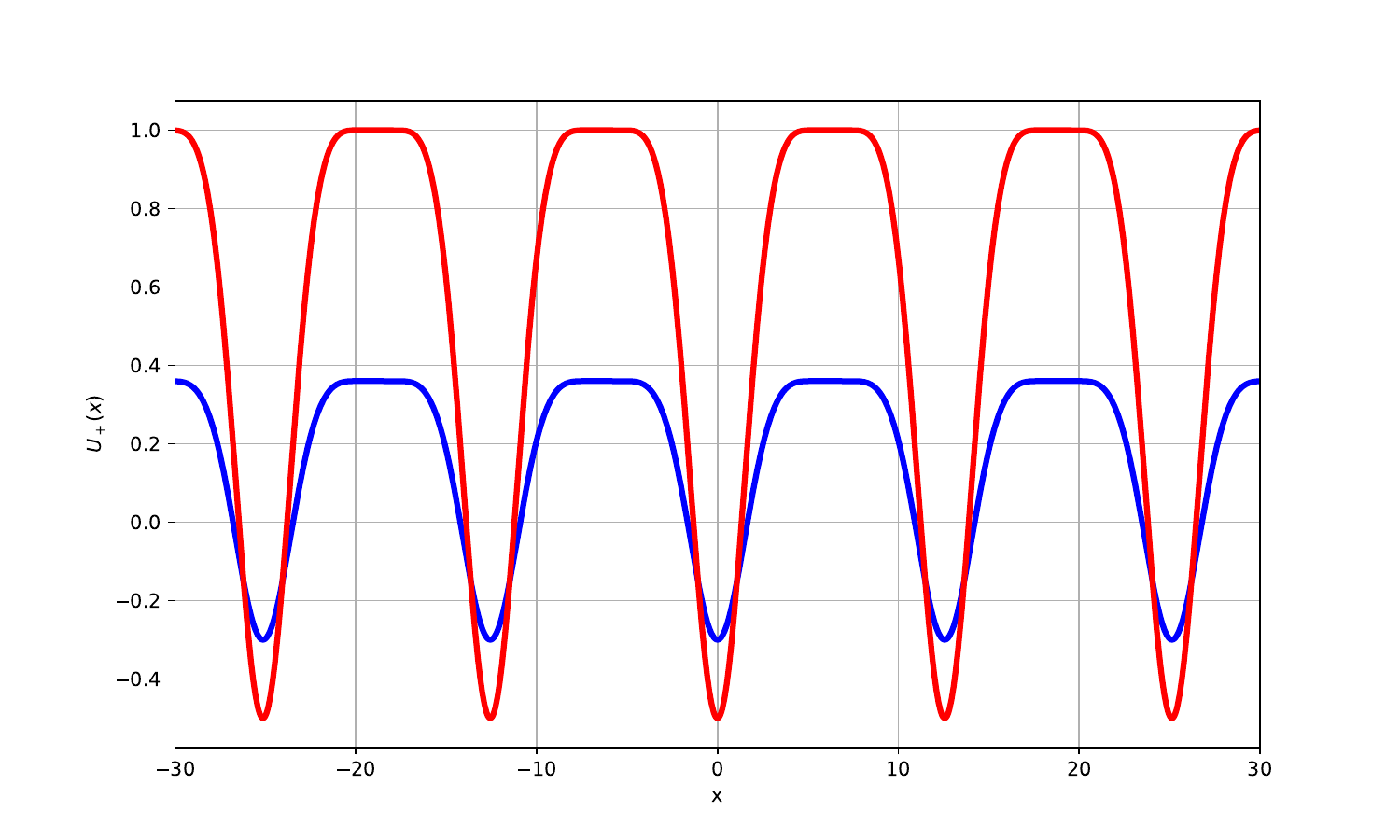}
    \caption{(Top panel) The potential $U_+(x)$ for $\alpha=0.3$ (blue) and $\alpha=0.5$ (red), with $g=0.4$. (Bottom panel) Same potential for $g=0.3$ (blue) and $g=0.5$ (red), with $\alpha=0.5$.}
  \label{fig1}
\end{center}
\end{figure}

After working with the Dirac equations \eqref{DE1} and \eqref{DE2}, we arrive at the decoupled Schrödinger-like equations

\begin{eqnarray}\label{schroeq}
       \bigg( - \frac{d^2}{dx^2} + U_{\pm}(x) \bigg) \psi_{\pm} & = & E^2\psi_{\pm},
\end{eqnarray}
where $U_\pm(x)$ can be identified as

\begin{eqnarray}
U_{\pm}(x) = 4g^2\phi^2 \mp 2g\frac{d\phi}{dx}.
    \label{potU}
\end{eqnarray}
Here, $\phi=\phi(x)$ is given by Eq. \eqref{phi2} and we substitute it into the potentials $U_{\pm}(x)$ to get to  

\begin{eqnarray}
U_{\pm}(x)\! =\! 4g^2\!\tanh^2(y(x))\! \mp\! g(1\!+\!\alpha^2 y^2(x))\!\sech^2(y(x)),\;\;\;
\end{eqnarray}
where $y(x)=(1/\alpha)\tan({\alpha x}/{2}) $. The potential $U_+(x)$ has the shape of a periodic potential which promotes the existence of bound states. It is depicted in Fig. \ref{fig1} for some values of $\alpha$ and $g$. 

It should be noted that the minimum of the potential $U_+(x)$ is at $-g$, with the maximum at $4g^2$. This maximum shows that the thresholds of energy for the fermion field that separate the presence of discrete bound states from the continuous spectrum are at $\pm2g$. Moreover, the variation of $\alpha$ causes the position of the wells to shift. This is related to the constant $d=2\pi/\alpha$, which represents the lattice parameter, such that $U_+(x+d)=U_+(x)$. We further notice that $\alpha$ does not affect the depth of the wells, which is clearly altered by changing the coupling parameter $g$.  For this reason, the emergence of bound states may be more noticeable for large values of $g$. We remind that the potential has at least one bound state, the zero mode, for distinct values of $\alpha$ and $g$.

\begin{table}[htbp]
  \centering
  \caption{Number of bound states $N$ for $g=0.3$, $0.5$ and $0.7$, for some values of $\alpha$.}
\bigskip
    \begin{tabular}{|c c c|c c c|c c c|}
    \hline
    \hline
    \quad $g$ \quad\quad & $\alpha$ \quad\quad & $N$ \quad & \quad $g$ \quad\quad & $\alpha$ \quad\quad & $N$ \quad & \quad $g$ \quad\quad & $\alpha$ \quad\quad & $N$ \quad \\
    \hline
    \hline
               \quad\quad & 0.1   \quad\quad & 3  \quad &             \quad\quad & 0.1   \quad\quad & 5  &            \quad\quad & 0.1   \quad\quad & 7\\
    \quad 0.3  \quad\quad & 0.3   \quad\quad & 7  \quad & \quad 0.5   \quad\quad & 0.3   \quad\quad & 13 & \quad 0.7  \quad\quad & 0.3   \quad\quad & 19\\
               \quad\quad & 0.5   \quad\quad & 11 \quad &             \quad\quad & 0.5   \quad\quad & 21 &            \quad\quad & 0.5   \quad\quad & 31\\
    \hline
    \end{tabular}%
  \label{tab}%
\end{table}%

\begin{figure}[!ht]
\begin{center}
  \centering
  {\includegraphics[width=0.45 \textwidth]{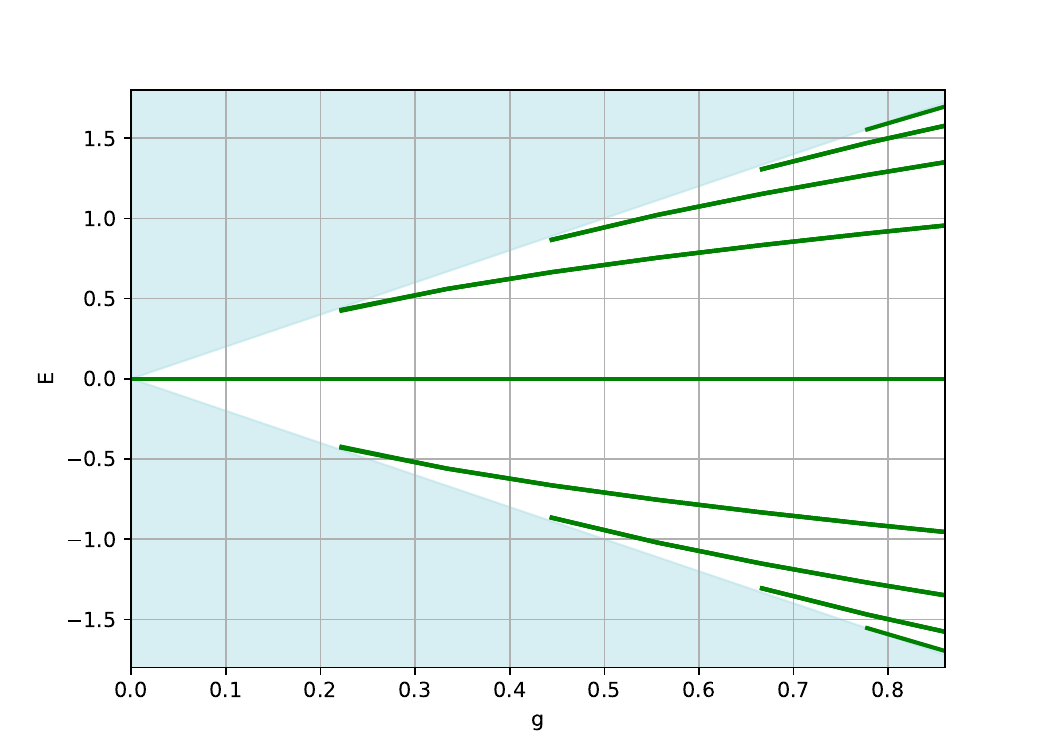}}    
  {\includegraphics[width=0.45 \textwidth]{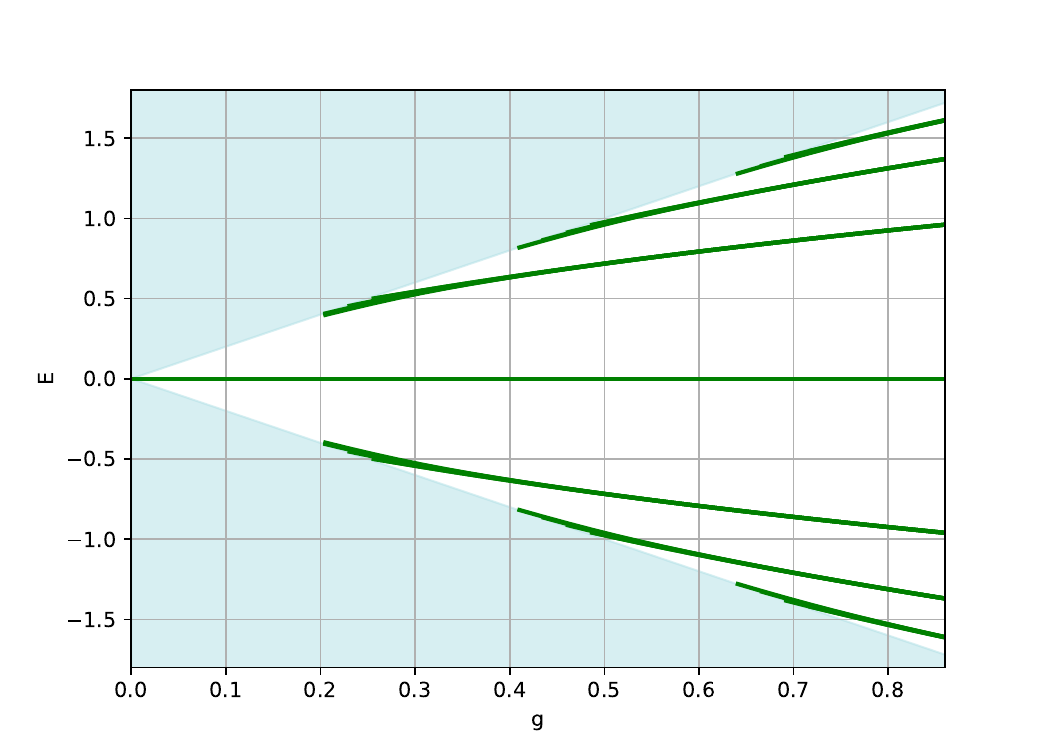}}
    \caption{Fermion spectrum as a function of $g$ for $\alpha=0.3$ (top) and for $\alpha=0.5$ (bottom).}
  \label{fig2}
\end{center}
\end{figure}

Let us now numerically compute the fermion spectrum. In order to do this, we use boundary conditions which are frequently employed in quantum mechanical problems. We consider the spatial coordinate $x$ in the interval $[-x_{max},x_{max}]$, with the spatial discretization $\delta x=0.02$. Since the potential is periodic, the spatial interval has to contain an integer number of the lattice parameter $d=2\pi/\alpha$, which depends on $\alpha$; for this reason, we use $x_{max}\approx 30$, meaning that it is as close to $30$ as possible. To solve the Eq. \eqref{schroeq}, we suppose that both the wave function and its derivative vanish at the edges of the spatial interval. Table \ref{tab} shows the number of bound states ($N$) for $g=0.3$, $0.5$, and $0.7$, and for $\alpha=0.1, 0.3,$ and $ 0.5$. We observe an increase in the number of bound states as one increases $\alpha$ and $g$. This is due to the fact that $\alpha$ controls the periodicity of the potential, and $g$ the depth of the wells, as illustrated in Fig. \ref{fig1}. The result motivates us to investigate the energy eigenvalues of the fermion bound states as one varies $g$. The energy spectrum as a function of $g$ for $\alpha=0.3$ and $\alpha=0.5$ are depicted in Fig. \ref{fig2}. In this figure, the bound states are restricted to appear in between the boundary lines, while the shaded area defines the region of scattering states. Since the system possesses energy conjugation symmetry, the spectrum is symmetric with respect to positive and negative values of $E$. We notice that, in addition to the zero mode, there are several other lines of negative and positive energies.

In particular, apart from the zero mode, every line in Fig.~\ref{fig2} represents a finite set of degenerate states. We have checked several possibilities and in this figure, for $\alpha=0.3$ and $g=0.8$ there are three overlapped lines of energies (threefold degeneracy) for each one of the eight lines of nonvanishing energies, and for $\alpha=0.5$ and $g=0.8$ there are five overlapped lines of energies (fivefold degeneracy) for each one of the six lines there depicted. In Ref.~\cite{baz1}, the emergence of new bound states within the fermionic gap was noticed. Here, however, the additional states are degenerated, which is due to the periodic nature of the potential under investigation. Consequently, the novel behavior revealed by the present study is associated with the emergence of degenerate states within the fermionic gap. As $\alpha$ increases, the wells of the potential approach one another, as shown in Fig. \ref{fig1}, enlarging the number of degenerate states for each of the nonvanishing energy eigenvalues. This shows that the degeneracy in the energy eigenvalues depends directly on $\alpha$, which controls the period of the potential, given by $2\pi/\alpha$. We have also fixed $g=0.7$ and $\alpha=0.5$, and increased the spatial interval $[-x_{max},x_{max}]$ to add new numerical results. For $x_{max}=70$ and $x_{max}=120$, we counted the number of bound states, obtaining $N=67$ and $N=115$, respectively. These results show that the degeneracy changes to $11$ or $19$, respectively, indicating that in the infinite lattice the bound states become bands of states.

As it is well known, the presence of localized structure is of direct importance to understand electric properties of organic polymers like polyacetylene \cite{Heeger}. Since trans-polyacetylene can have two degenerate ground states, it may develop soliton-like localized structure \cite{jackiw1,Heeger} which directly contributes to the transport of electric charge as they can incorporate dopants that donate or accept electrons, inducing the transport of electric charges. If a periodic array of localized structures is added to the polymeric chain, the results of the present work may suggest important modification in the electronic transport, highlighting the possibility to add new strategies to increase the conductivity of organic polymers \cite{con1,con2}.

\begin{figure}[!ht]
\begin{center}
  \centering
   {\includegraphics[width=0.12 \textwidth]{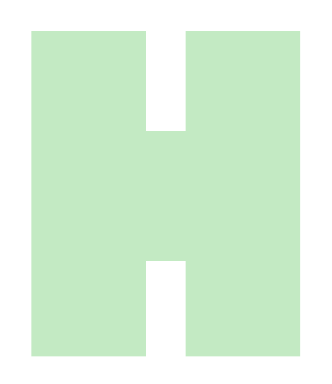}}
   {\includegraphics[width=0.48 \textwidth]{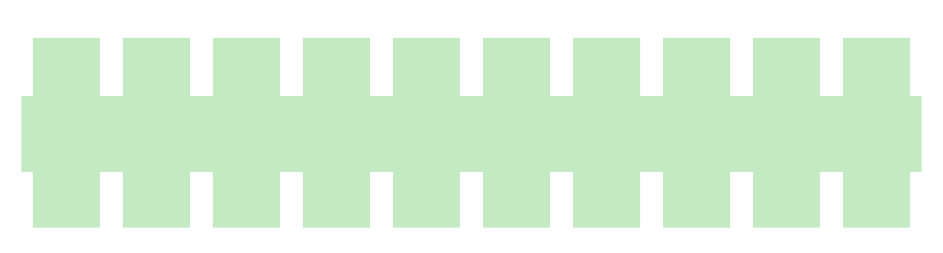}}      
   \caption{Illustration of the H-shaped magnetic structure used in \cite{IBM} (top) and a short piece of a nanoribbon of Ni$_{70}$Fe$_{30}$/Fe (bottom) similar to the one considered in \cite{PRL2008}, which may support a periodic array of kinks.}
  \label{fig3}
\end{center}
\end{figure}
\begin{figure}[!ht]
\begin{center}
  \centering
   {{\includegraphics[width=0.48 \textwidth]{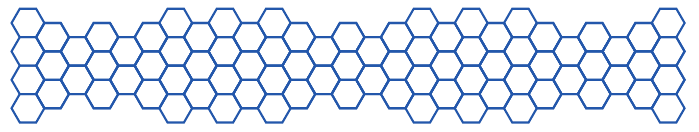}}}  
   \caption{Part of a graphene nanoribbon which may be engineered to behave as a one dimensional nanowire with controllable conductivity.}
  \label{fig4}
\end{center}
\end{figure}

Since we are working in one spatial dimension, we can also think on the possibility to consider the above results to investigate the electronic transport in a long and almost one-dimensional magnetic ribbon at the nanometric scale. If one takes a nanoribbon of Ni$_{70}$Fe$_{30}$/Fe similar to the one used in \cite{PRL2008}, we believe it is possible to build a periodic structure as illustrated in the bottom panel in Fig. \ref{fig3}. This construction is inspired in the result presented before in Ref. \cite{IBM}, where a geometric constriction similar to the one depicted in the top panel in Fig. \ref{fig3} at the nanometric scale was able to modify importantly the structure of the magnetization within the geometric constriction, and also in Ref. \cite{baz1}, where the geometric modification was shown to induce new bound states inside the fermionic gap. In this sense, the periodicity of the magnetic structure displayed in Fig. \ref{fig3} (bottom) may induce degeneracy of the fermionic states and modify the transport of charge within the magnetic ribbon, an effect that could be experimentally measured under conditions similar to the ones considered in Ref. \cite{PRL2008}. This is a feasible proposal to construct magnetic ribbons at the nanometric scale with controllable conductivity.  Another possibility of current interest concerns the study of graphene nanoribbons, which are also almost one-dimensional materials with a graphitic lattice structure \cite{natrev}. If one works to add a periodic array of localized structures, it may also act as a linear crystal of topological defects and induce important modifications in the transport of electronic current. A suggestion here is the engineering of one dimensional graphene nanoribbons superlattices as the ones described in Refs. \cite{eng,new}. An illustration is displayed in Fig. \ref{fig4}.

The degeneracy unveiled in the present study may also be of interest when considering fermions in one-dimensional optical lattices \cite{OL} and in other scenarios, in particular, when one adds another spatial dimension, working with planar systems, thinking of fermions in a periodic array of localized structures, as in the case of twisted bilayer graphene, with the formation of regular moiré patterns \cite{TBG,moire}.

\section*{Acknowledgments}

We would like to thank Matheus Marques for discussions. This work was partially supported by Conselho Nacional de Desenvolvimento Científico e Tecnológico (CNPq, Grants No. 303469/2019-6 (DB) and No. 402830/2023-7 (DB)), by Coordenação de Aperfeiçoamento de Pessoal de Nível Superior (CAPES, Finance Code 001), by Fundação de Amparo à Pesquisa e ao Desenvolvimento Científico e Tecnológico do Maranhão (FAPEMA, Grant no. 07838/17 (FCS)), and by Paraiba State Research Foundation (Grant no. 0015/2019 (DB)).


\end{document}